\documentclass[showpacs,twocolumn,nofootinbib]{revtex4}
\usepackage{amsfonts}
\usepackage{graphicx}
\usepackage{amsmath}
\usepackage{amssymb}
\usepackage{amsthm}
\usepackage{verbatim}

\begin{document}

\title{Criticality and Big Brake singularities in the tachyonic evolutions
of closed Friedmann universes with cold dark matter}
\author{Zsolt Horv\'{a}th$^{1}$, Zolt\'{a}n Keresztes$^{1,2}$, Alexander Yu.
Kamenshchik$^{3,4}$, L\'{a}szl\'{o} \'{A}. Gergely$^{1,2}$}
\affiliation{$^{1}$ Department of Theoretical Physics, University of Szeged, Tisza Lajos
krt 84-86, Szeged 6720, Hungary \\
$^{2}$ Department of Experimental Physics, University of Szeged, D\'{o}m T%
\'{e}r 9, Szeged 6720, Hungary \\
$^{3}$ Dipartimento di Fisica e Astronomia, Universit\`{a} di Bologna and
INFN, via Irnerio 46, 40126 Bologna, Italy \\
$^{4}$ L.D. Landau Institute for Theoretical Physics, Russian Academy of
Sciences, Kosygin street 2, 119334 Moscow, Russia}

\begin{abstract}
The evolution of a closed Friedmann universe filled by a tachyon scalar
field with a trigonometric potential and cold dark matter (CDM) is
investigated. A subset of the evolutions consistent to 1$\sigma $ confidence
level with the Union 2.1 supernova data set is identified. The evolutions of
the tachyon field are classified. Some of them evolve into a de Sitter
attractor, while others proceed through a pseudo-tachyonic regime into a
sudden future singularity. Critical evolutions leading to Big Brake
singularities in the presence of CDM are found and a new type of
cosmological evolution characterized by singularity avoidance in the
pseudo-tachyon regime is presented.
\end{abstract}

\maketitle

\section{Introduction}

The discovery of the cosmic acceleration \cite{cosmic} has generated a lot
of interest in Friedmann cosmological models with various kinds of exotic
forms of matter, collectively known as dark energy \cite{dark}.
Unsurprisingly, some of these models allow for novel type of singularities
(occurring in finite time), which can be either strong or weak \cite{Tipler}%
, \cite{Krolak}, \cite{cross}, \cite{Cosmofun}.

In particular, the Big Rip singularity occurs at infinite scale factor,
diverging Hubble parameter $H$ and diverging $\dot{H}$ \cite{BigRip}, \cite%
{classification}. This singularity is strong according to both Tipler's \cite%
{Tipler} and Kr\'{o}lak's \cite{Krolak} definition. Some other singularities
are alike, but they occur at finite scale factor (FSF). The FSF singularity 
\cite{classification}, \cite{FSF} is also strong according to Kr\'{o}lak's
definition, but is classified as weak according to Tipler's definition.

The sudden future singularity (SFS), discovered many years ago \cite%
{Barrow86}, is characterized by both a finite scale factor and a finite
Hubble parameter, while only $\dot{H}$ diverges \cite{SFS}. As consequence
the energy density is finite, while the pressure diverges at SFS. The
geodesic equation (containing only $H$) remains regular at SFS, such that
point particles will pass through the singularity. However for an
infinitesimally short time the tidal forces in the geodesic deviation
equation become infinite due to the diverging $\dot{H}$. The SFS singularity
is weak in both Tipler's and Kr\'{o}lak's definitions, hence it has been
conjectured that finite object may also cross this singularity \cite{cross}.

A subclass of the SFS singularities is the Big Brake, introduced in Ref. 
\cite{tach1} as emerging in the presence of a particular tachyonic scalar
field (i.e. Born-Infeld-type field). Such a singularity is characterized by
a full stop in the expansion, occurring at finite scale factor and vanishing
energy density, augmented by a diverging deceleration and pressure. There is
also the w-singularity \cite{classification}, \cite{w}, when both the
pressure and energy density vanish, however their ratio, the barotropic
index $w$ diverges, is also weak in both definitions.

Interesting features of the SFS singularities have been found on the example
of the flat Friedmann universe filled with a combination of cold dark matter
(dust) and tachyonic field, the latter obeying a particular potential of
trigonometric functions \cite{tach1}. The model possesses two surprising
features. First, for certain model parameter range, at some stage of the
cosmological evolution, the tachyonic field is constrained to transform into
another Born-Infeld-type field, the pseudo-tachyon. Secondly, in this regime
the universe unavoidably evolves into the SFS.

In this model in the absence of the dust component the SFS singularity
represents a Big Brake. Confrontation with type Ia Supernovae (SNIa) data
showed that the allowed cosmological evolutions can reach either the Big
Brake or the de Sitter attractor \cite{tach2}. Further, in Ref. \cite{tach3}
it was proven that the Big Brake singularity is traversable, and the
evolution continues through a collapsing universe into a Big Crunch.

When the dust component is included, the SFS cannot be a Big Brake anymore 
\cite{tach4}. Indeed, the universe reaches the singularity with a finite but
nonzero $H$. Its traversability is guaranteed by certain changes in the
matter properties \cite{change}, \cite{tach5}, imposed by the requirement of
a smooth passage through the SFS. These changes lead to the notion of the
quasi-tachyonic Born-Infeld-type field, a phantom field \cite{Caldwell} with
negative energy density. A throughout confrontation of these models with
SNIa, baryonic acoustic peaks, and cosmic microwave background temperature
anisotropy power spectrum did not disrule the evolutions leading to either
the de Sitter attractor or to the SFS \cite{tach6}.

The present work represents a continuation of the series of papers \cite%
{tach1}, \cite{tach2}--\cite{tach5}, \cite{tach6}, investigating the
consequences of the introduction of a tiny positive curvature, which remains
compatible with the recent Planck data \cite{Planck2013}, \cite{Planck2015}.
As will be shown, the addition of the curvature term enriches the velocity
phase space of the possible cosmological evolutions, adding to it regimes
which were impossible to realize in the flat model. We study these regimes
numerically and show the existence of a critical behavior in the formation
of the singularities in the framework of these tachyonic cosmological
models. Such critical phenomena were well-known in the numerical study of
gravitational collapse of a scalar field \cite{Choptuik} and of various
other fields (as reviewed in Ref. \cite{LivRevCriticalCollapse}), with the
critical evolutions separating black-hole reaching evolutions from those
which do not lead to black hole formation.

The structure of the paper is as follows. In Section \ref{mix} we introduce
the equations of the tachyonic cosmology in the presence of dust and
curvature. In Section \ref{SNIatest} we confront the luminosity
distance-redshift relation with the supernova data. In Section \ref%
{evolution} we investigate numerically the dynamics and give the
classification of all possible regimes, with emphasis on the new types of
evolutions. The results are summarized in the Conclusions.

The units are chosen as $c=1$ and $8\pi G/3=1$.

\section{A mixture of dust and tachyonic field in a Friedmann universe \label%
{mix}}

We consider a Friedmann universe 
\begin{equation}
ds^{2}=dt^{2}-a^{2}\left( t\right) \left( \frac{1}{1-Kr^{2}}%
dr^{2}+r^{2}d\theta ^{2}+r^{2}\sin ^{2}\theta d\phi ^{2}\right) ~,
\end{equation}%
with cosmological time $t$, comoving radial distance $r$, polar and
azimuthal angles $\theta $ and $\phi $, unnormalized curvature index $K>0,=$ 
$0$ or $<0$ and scale factor $a(t)$. The combined energy momentum tensor is
an ideal fluid with energy density $\rho $ and pressure $p$. The dynamics of
the scale factor is governed by the Raychaudhuri equation: 
\begin{equation}
\dot{H}-\frac{K}{a^{2}}+\frac{3}{2}\left( \rho +p\right) =0~,  \label{Raych}
\end{equation}%
and the change in the energy density by the continuity equation: 
\begin{equation}
\dot{\rho}+3H\left( \rho +p\right) =0~.  \label{cont}
\end{equation}%
The overdot denotes time derivative and $H\equiv \dot{a}/a$ is the Hubble
parameter. The first integral of the system (\ref{Raych})-(\ref{cont}) is
the Friedmann equation:%
\begin{equation}
H^{2}+\frac{K}{a^{2}}=\rho ~.  \label{Friedmann}
\end{equation}%
We assume that the Friedmann universe is filled with a mixture of tachyonic
field and cold dark matter (CDM). The mixture of the two ideal fluid
components (distinguished by $m$ and $T$ subscripts) is characterized by $%
\rho =\rho _{m}+\rho _{T}$ and $p=p_{T}$, where 
\begin{equation}
\rho _{m}=\frac{\rho _{m}^{\ast }}{a^{3}}~,~  \label{dust}
\end{equation}%
with integration constant $\rho _{m}^{\ast }$. The evolution of the
homogeneous tachyonic field $T(t)$ follows from the Lagrangian density \cite%
{Sen}%
\begin{equation}
L_{T}=-\sqrt{-g}V\left( T\right) \sqrt{1-g^{tt}s^{2}}~,  \label{tachLagreal}
\end{equation}%
with $s=\dot{T}$. Variation with respect to the metric gives its energy
density%
\begin{equation}
\rho _{T}=\frac{V\left( T\right) }{\sqrt{1-s^{2}}}~,  \label{realdens}
\end{equation}%
and pressure%
\begin{equation}
p_{T}=-V\left( T\right) \sqrt{1-s^{2}}~.  \label{realpres}
\end{equation}%
As $w=p/\rho =s^{2}-1$, in principle the tachyonic field can mimic dark
energy. The particular case of a trigonometric tachyonic potential \cite%
{tach1} 
\begin{equation}
V=\frac{\Lambda \sqrt{1-(1+k)y^{2}}}{1-y^{2}}~,  \label{realpot}
\end{equation}%
with%
\begin{equation}
y=\cos \left[ \frac{3}{2}{\sqrt{\Lambda \,(1+k)}T}\right] ~,  \label{y-def}
\end{equation}%
$\Lambda >0$ and $-1<k<1$ model parameters further leads to interesting
behavior, including a smooth passage into the $s>1$ regime and a sudden
future singularity which can be passed through by geodesics which then
bounce back into a recollapse \cite{tach1}--\cite{tach6}.

Note that the system is invariant under the simultaneous changes $\left(
y\rightarrow -y,~s\rightarrow -s\right) $.This yields a double coverage of
the dynamics in these velocity phase space variables.

\section{Test with supernovae data \label{SNIatest}}

\begin{figure}[th]
\includegraphics[height=8.5cm,angle=270]{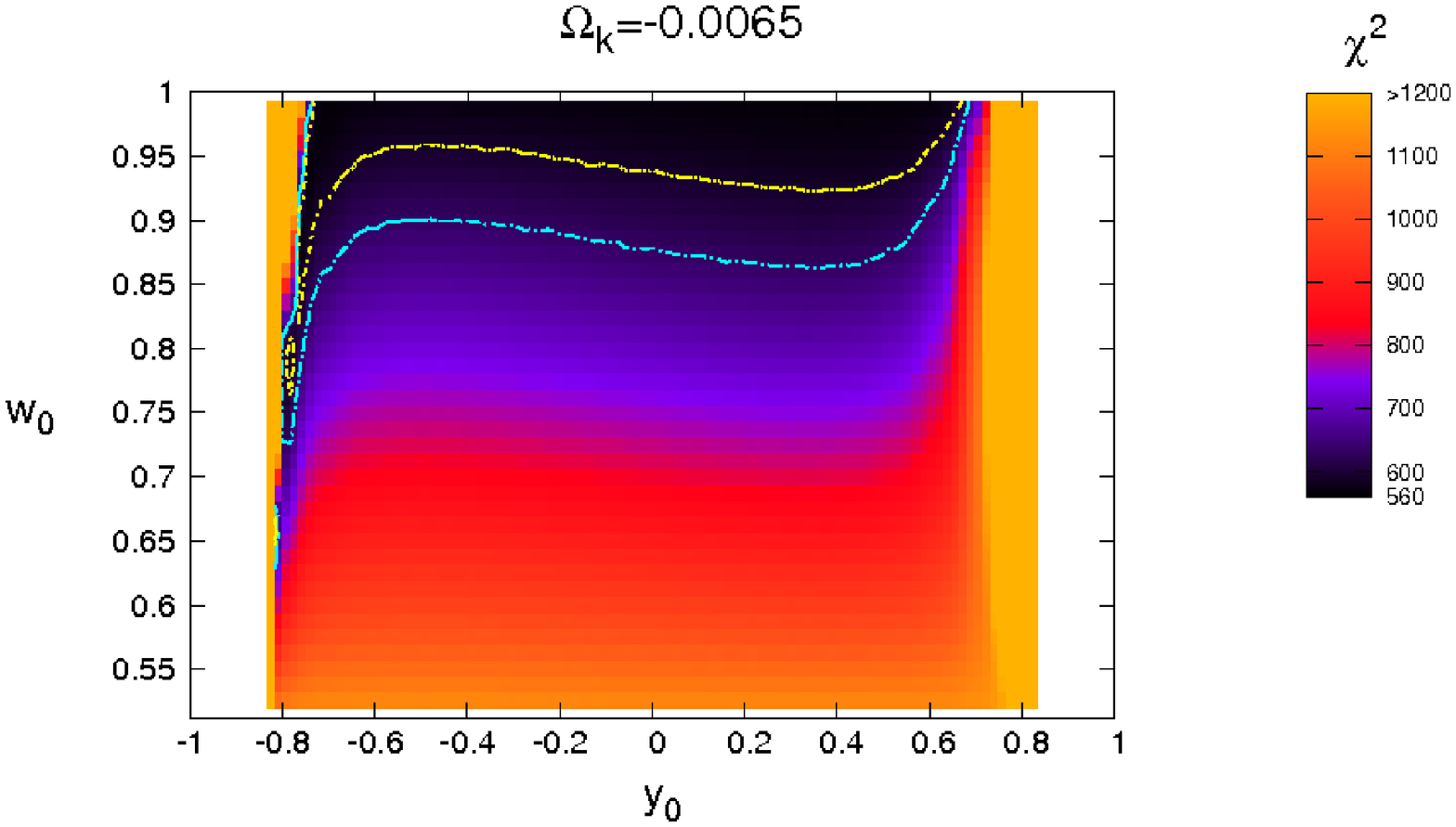} %
\includegraphics[height=8.5cm, angle=270]{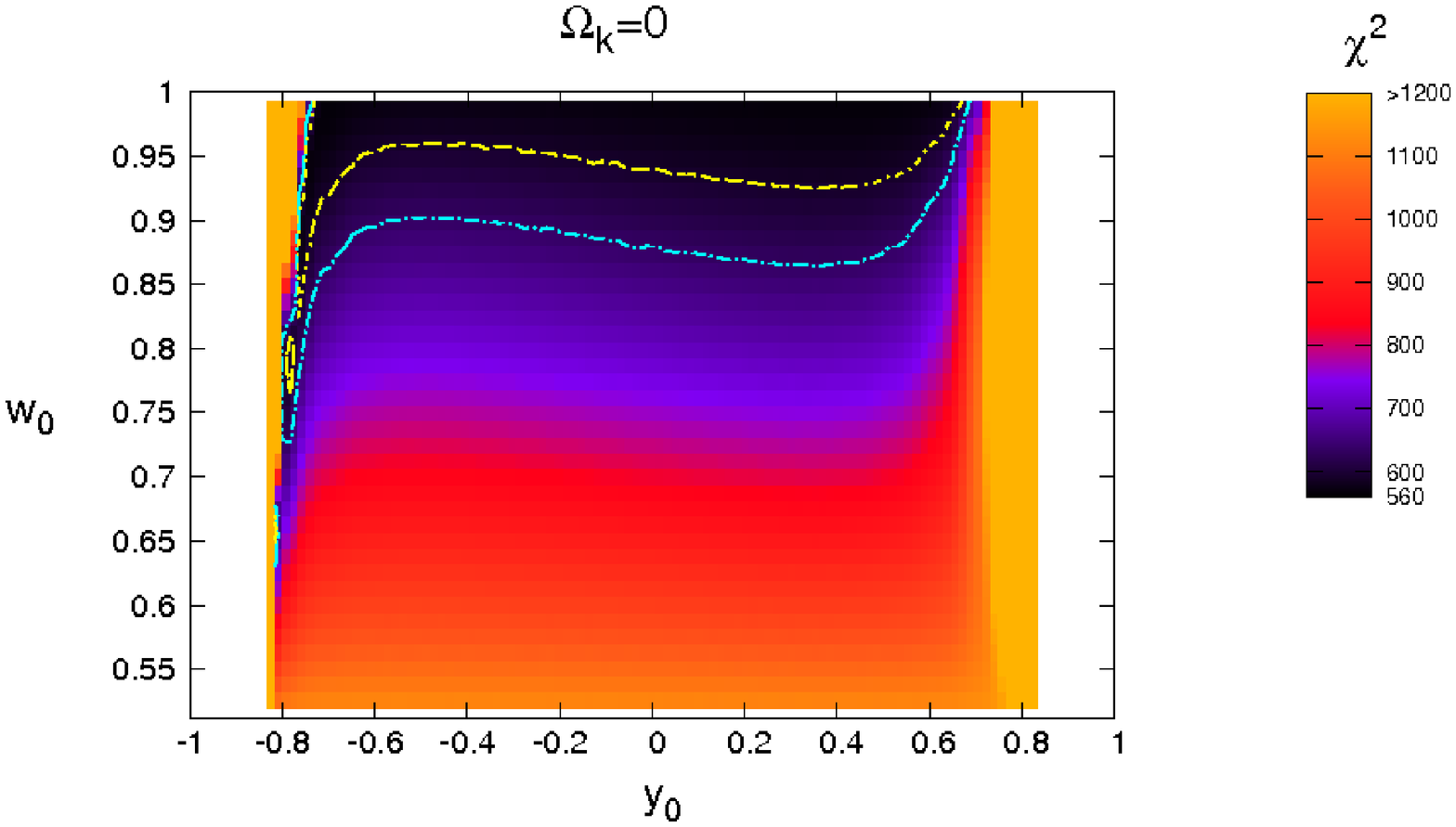}
\caption{(Color online) The fit of the luminosity distance vs. redshift for
the parameters $k=0.44$ and $\Omega _{m}=0.315$ in the parameter plane ($%
y_{0}$, $w_{0}=\left( 1+s_{0}^{2}\right) ^{-1}$) (upper panel for $\Omega
_{K}=-0.0065$, lower panel for $\Omega _{K}=0$). The contours refer to the $1%
\protect\sigma $ and $2\protect\sigma $ confidence levels. The colour code
of $\protect\chi ^{2}$ is indicated on the vertical stripes.}
\label{contur}
\end{figure}

The evolution of the Friedmann universe is determined by the initial
conditions $H_{0}=H\left( z=0\right) $, $y_{0}=y\left( z=0\right) $, $%
s_{0}=s\left( z=0\right) $, ($z$ is the redshift), the present values of the
cosmological parameters 
\begin{equation}
\Omega _{m}=\frac{\rho _{m}^{\ast }}{a_{0}^{3}H_{0}^{2}}~,~\Omega _{K}=-%
\frac{K}{a_{0}^{2}H_{0}^{2}}~,  \label{new-var}
\end{equation}%
associated to the CDM and the curvature, and the model parameters $\Lambda $%
, $k$. The parameter $\Omega _{K}$ is constrained by observations to 95\%
confidence level: $-0.0065\leq \Omega _{K}\leq 0.0012$ by WMAP data \cite%
{WMAP12} and $-0.019\leq \Omega _{K}\leq 0.011$ by Planck data (see fifth
column in Table 5 of \cite{Planck2015}). In what follows we set $\Omega
_{m}=0.315$ (supported by the cosmic microwave anisotropies measured by the
Planck satellite \cite{Planck2013}, \cite{Planck2015}), $k=0.44$, $H_{0}=70$
km/sec/Mpc and we analyze both $\Omega _{K}=-0.0065$ and $0$ (the former
being compatible with $K>0$, the latter representing a spatially flat
universe). The Friedmann equation (\ref{Friedmann}) represents a constraint
among $y_{0}$, $s_{0}$ and $\Lambda $, thus only the first two are varied.
The confrontation of the tachyonic model with the Union 2.1 SNIa data set 
\cite{Union21} through a $\chi ^{2}$-test as described in Ref. \cite{tach2}
relies on fitting with the computed luminosity distance ($d_{L}$)-redshift
relation. A dimensionless luminosity distance $\hat{d}_{L}=H_{0}d_{L}$ is
given by 
\begin{eqnarray}
\sqrt{-\Omega _{K}}\frac{\hat{d}_{L}}{1+z} &=&\chi \left( z\right) \text{%
~for }~K=0~, \\
\sqrt{-\Omega _{K}}\frac{\hat{d}_{L}}{1+z} &=&sin\left( \chi \left( z\right)
\right) ~\text{for}~K>0~,
\end{eqnarray}%
with 
\begin{equation}
\frac{d\chi \left( z\right) }{dz}=\frac{\sqrt{-\Omega _{K}}}{\hat{H}}~.
\end{equation}%
($\hat{H}=H/H_{0}$) We show the fit on Fig \ref{contur}. The contours refer
to the $68.3\%$ (1$\sigma $) and $95.4\%$ (2$\sigma $) confidence levels.

Thus the introduction of the curvature does not hamper the compatibility of
the model with the SNIa observations.

\section{Velocity phase space \label{evolution}}

The $y^{2}<1/\left( 1+k\right) $ , $s^{2}<1$ domain of the velocity phase
space represents proper tachyonic fields, characterized by the Lagrangian (%
\ref{tachLagreal}), energy density (\ref{realdens}), pressure (\ref{realpres}%
) and potential (\ref{realpot}). A subset of the initial conditions $y_{0}$, 
$s_{0}$ allow the universe to evolve outside this regime through the corner
points $y,s=\pm 1$ where the potential $V$ vanishes, as described in Ref. 
\cite{tach1}. These are exceptional points, where the curvature does not
diverge when $s^{2}=1$. When $s^{2}>1,\left( 1+k\right) ^{-1}<y^{2}<1$ the
field becomes pseudo-tachyonic and $V$ imaginary. Although the Lagrangian (%
\ref{tachLagreal}) does not change in this regime, it is convenient to
introduce the real potential 
\begin{equation}
W\equiv iV=\frac{\Lambda \sqrt{\left( 1+k\right) y^{2}-1}}{1-y^{2}}~.
\end{equation}%
Note that in the Lagrangian $W$ absorbs the imaginary factor arising from
the change of sign in the equation of state $w=s^{2}-1$. The expressions of
the energy density and pressure are also unchanged in the pseudo-tachyonic
regime. Rewriting the Lagrangian, the energy density and pressure in terms
of $W$ however manifestly shows that all these quantities stay real: 
\begin{eqnarray}
L_{P} &=&\sqrt{-g}W\sqrt{g^{tt}s^{2}-1}~, \\
\rho _{P} &=&\frac{W\left( T\right) }{\sqrt{s^{2}-1}},  \label{rhoP} \\
p_{P} &=&W\left( T\right) \sqrt{s^{2}-1}~.
\end{eqnarray}%
The pressure is positive, thus the expansion of the universe slows down in
this regime.

Further evolution to $s\rightarrow \pm \infty $ leads to SFS, arising for
finite values of $z$, $\hat{H}$ and $W$ and $\rho _{P}\rightarrow 0$, $%
p_{P}\rightarrow \infty $. As was shown in Ref. \cite{tach5} the evolution
can be continued smoothly across these soft singularities at the price of a
sudden change of the Lagrangian, which will represent a quasi-tachyonic
field:%
\begin{eqnarray}
L_{Q} &=&\sqrt{-g}W\sqrt{g^{tt}s^{2}+1}~,  \label{rhoQ} \\
\rho _{Q} &=&\frac{-W\left( T\right) }{\sqrt{s^{2}+1}}, \\
p_{Q} &=&W\left( T\right) \sqrt{s^{2}+1}~.
\end{eqnarray}%
Note that $w_{Q}=p_{Q}/\rho _{Q}=-\left( s^{2}+1\right) \neq w$, thus the
quasi-tachyon is a phantom with negative energy density.

In order to formulate the dynamical equations with dimensionless variables,
we normalize the potentials as 
\begin{eqnarray}
\hat{V} &=&\frac{\Omega _{\Lambda }\sqrt{1-(1+k)y^{2}}}{1-y^{2}}~, \\
\hat{W} &=&\frac{\Omega _{\Lambda }\sqrt{\left( 1+k\right) y^{2}-1}}{1-y^{2}}%
~.
\end{eqnarray}%
(i.e. $\Omega _{\Lambda }=\Lambda /H_{0}^{2}$, $\hat{V}=V/H_{0}^{2}$, $\hat{W%
}=W/H_{0}^{2}$).

The dimensionless variables $z(t)$ and $\hat{H}(t)$ evolve as 
\begin{eqnarray}
{\frac{dz}{d\hat{t}}}+\left( 1+z\right) \hat{H} &=&0~,  \label{dzeq} \\
{\frac{d\hat{H}}{d\hat{t}}}+\frac{3}{2}\left( \hat{\rho}_{m}+\hat{\rho}+\hat{%
p}\right) +\Omega _{K}\left( 1+z\right) ^{2} &=&0~,  \label{RaychNorm}
\end{eqnarray}%
with $\hat{t}=H_{0}t$. Eq. (\ref{dzeq}) follows from the definition of $z$,
while Eq. (\ref{RaychNorm}) is the Raychaudhuri equation. Here $\hat{\rho}%
_{m}=\Omega _{m}\left( 1+z\right) ^{3}$ is the normalized density of the
dust, while $\hat{\rho}$ stands for either of the normalized densities 
\begin{eqnarray}
\hat{\rho}_{T} &=&\frac{\hat{V}}{\sqrt{1-s^{2}}}~, \\
\hat{\rho}_{P} &=&\frac{\hat{W}}{\sqrt{s^{2}-1}}~, \\
\hat{\rho}_{Q} &=&\frac{-\hat{W}}{\sqrt{s^{2}+1}}~,
\end{eqnarray}%
holding for the tachyonic, pseudo-tachyonic or quasi-tachyonic regimes,
respectively. Similarly, $\hat{p}$ denotes one of the normalized pressures 
\begin{eqnarray}
\hat{p}_{T} &=&-\hat{V}\sqrt{1-s^{2}}~, \\
\hat{p}_{P} &=&\hat{W}\sqrt{s^{2}-1}~, \\
\hat{p}_{Q} &=&\hat{W}\sqrt{s^{2}+1}~.
\end{eqnarray}

\begin{figure}[tbp]
\includegraphics[width=6cm, angle=270]{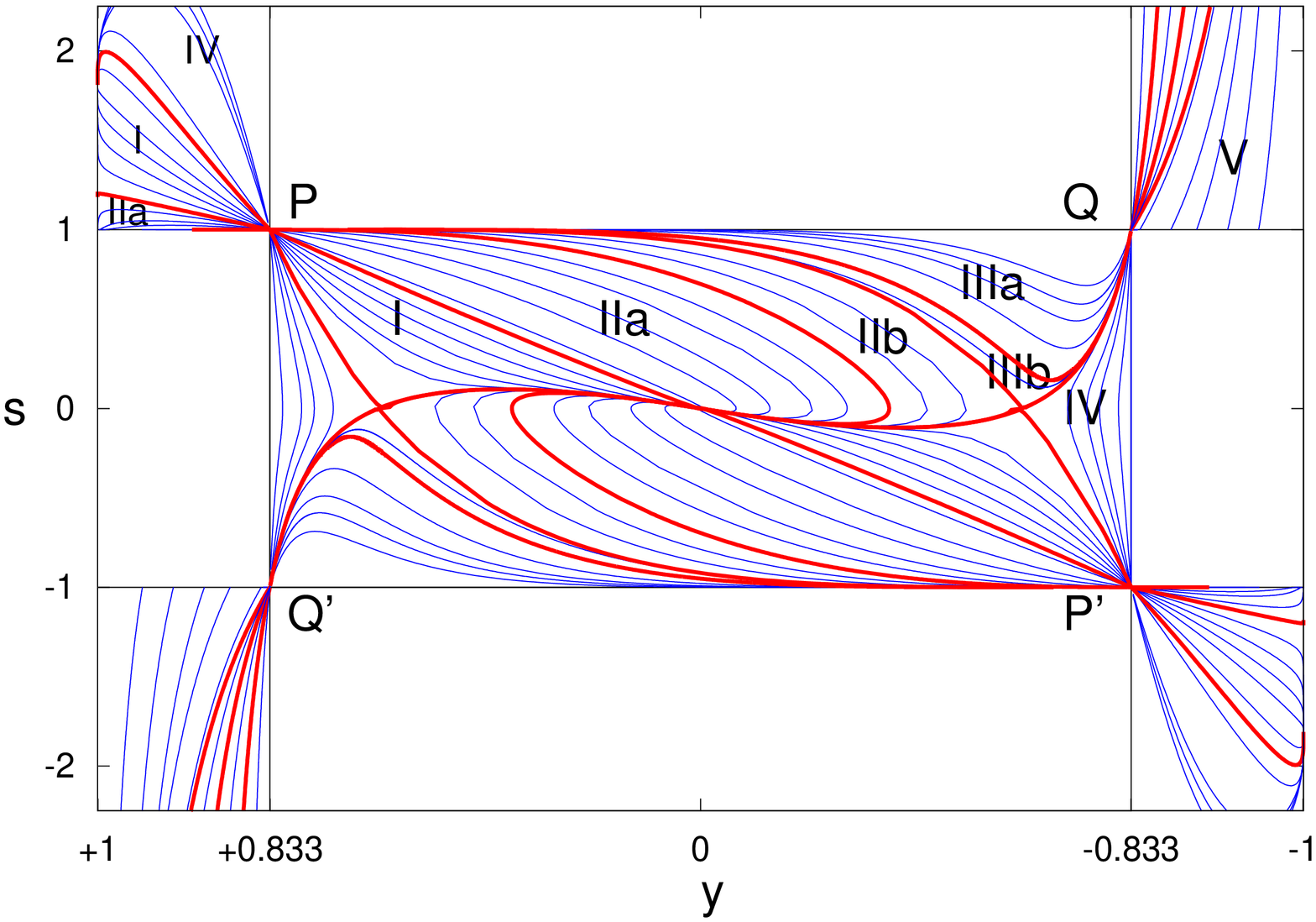}
\par
\includegraphics[width=6cm, angle=270]{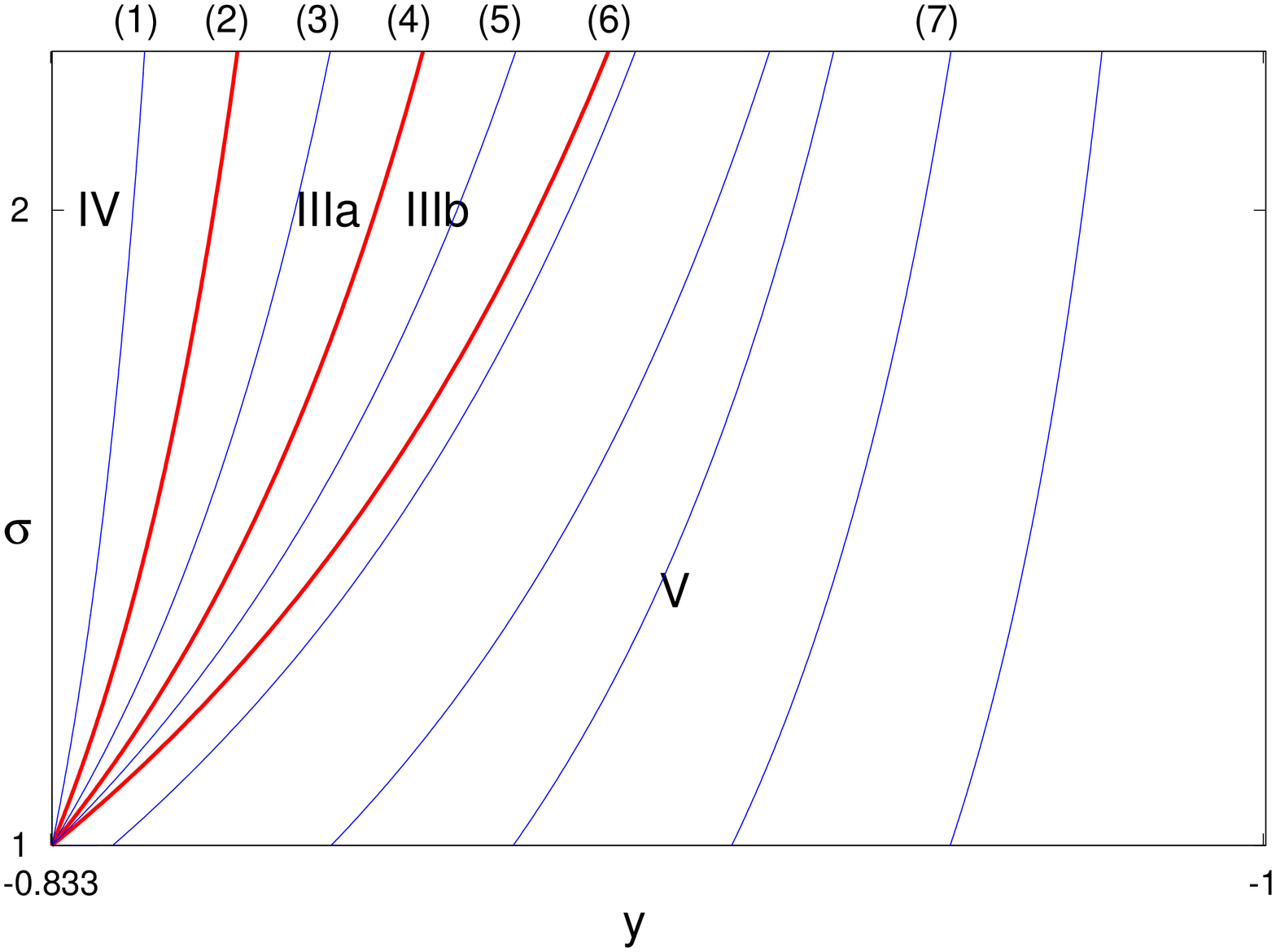}
\caption{(Color online) Upper panel: The numerical evolution of the
dynamical equations yields seven type of trajectories (I, IIa, IIb, IIIa,
IIIb, IV and V), separated by separatrices (red). Lower panel: the
pseudo-tachyonic evolutions of the upper right corner. We chose $\Omega
_{\Lambda }=0.8$.}
\label{fazisabra}
\end{figure}

The dynamics of the tachyonic field $y(t)$ emerges from the definition of $s$
as 
\begin{equation}
{\frac{dy}{d\hat{t}}}+\frac{3}{2}s\sqrt{{\Omega _{\Lambda }}\left(
1+k\right) \left( 1-y^{2}\right) }=0~,  \label{dydt}
\end{equation}%
while the evolution of $s(t)$ in the tachyonic, pseudo-tachyonic and
quasi-tachyonic regimes are given by the following Euler-Lagrange equations: 
\begin{eqnarray}
\frac{ds}{d\hat{t}}+\left( 1-s^{2}\right) \left( 3\hat{H}s+\frac{\hat{V}_{,T}%
}{\hat{V}}\right) &=&0~,  \label{pds} \\
\frac{ds}{d\hat{t}}+\left( 1-s^{2}\right) \left( 3\hat{H}s+\frac{\hat{W}_{,T}%
}{\hat{W}}\right) &=&0~, \\
\frac{ds}{d\hat{t}}+\left( 1+s^{2}\right) \left( 3\hat{H}s-\frac{\hat{W}_{,T}%
}{\hat{W}}\right) &=&0~.  \label{qds}
\end{eqnarray}%
In the neighborhood of the soft singularity $y$ can be approximated as 
\begin{eqnarray}
y_{P} &=&{{y_{S}}}+\frac{\sqrt{6}}{2}\sqrt{1-{{{y_{S}}}}^{2}}  \notag \\
&&\times \sqrt{{\Omega _{\Lambda }}\left( 1+k\right) }\sqrt{\frac{1}{H_{S}}}%
\sqrt{t_{S}-t}~,
\end{eqnarray}%
\begin{eqnarray}
y_{Q} &=&{{y_{S}}}-\frac{\sqrt{6}}{2}\sqrt{1-{{{y_{S}}}}^{2}}  \notag \\
&&\times \sqrt{{\Omega _{\Lambda }}\left( 1+k\right) }\sqrt{\frac{1}{H_{S}}}%
\sqrt{t-t_{S}}~.
\end{eqnarray}%
(These formulas are analogous to Eqs. (30) and (50) of Ref. \cite{tach5}
presented in terms of the original tachyonic variable $T$, and they hold for
the upper right strip of Fig. \ref{fazisabra}, where $y_{P}>y_{S}>y_{Q}$.)
The two expressions coincide at the singularity. In the quasi-tachyonic
regime the Hubble variable decreases, reaching zero at some point which is
followed by a recollapse through the same type of singularity to the
pseudo-tachyonic regime, ending in a Big Crunch. This was always the case
for $K=0$ \cite{tach3}.

Due to the extra gravitational attraction represented by $\Omega _{K}<0$
however some of the pseudo-tachyonic evolutions do not reach the
singularity, avoiding the quasi-tachyonic regime. Instead the recollapse
into a Big Crunch commences earlier. These two types of evolutions are
separated by a critical trajectory when the soft singularity is reached
exactly with $\hat{H}=0$. In what follows we will investigate this
criticality. 
\begin{figure}[tbp]
\hskip 0.5cm \includegraphics[width=6cm, angle=270]{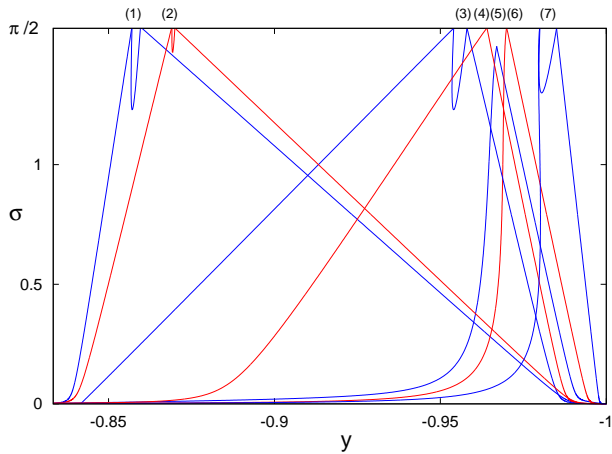} 
\caption{(Color online) The trajectories presented on the lower panel of
Fig. \protect\ref{fazisabra} are extended to infinity, and mapped into the
interval $[0,\protect\pi /2]$ by a conformal transformation $\protect\sigma %
=arctan(10^{-3}s)$. The pseudo-tachyonic and quasi-tachyonic regimes of each
of type IV, IIIa, IIIb and V evolutions (blue), and the separatrices between
types IV-IIIa, IIIa-IIIb and IIIb-V (red) are shown. The soft singularities
correspond to $\protect\sigma =\protect\pi /2$. The evolutions labeled by
numbers are identical to those represented on the lower panel of Fig. 
\protect\ref{fazisabra}. The critical evolutions leading to Big Brake
singularities in the presence of dust are (4) and (6).}
\label{pszeudokvazi}
\end{figure}

The numerical evolution of the dynamical equations yields seven types of
trajectories, I, IIa, IIb, IIIa, IIIb, IV and V, represented on the upper
panel of Fig. \ref{fazisabra}. The lower panel magnifies out the
pseudo-tachyonic evolutions of the upper right corner.

The trajectories of type I originate from the Big Bangs $(y,s)=(1,\sqrt{1+1/k%
})$ and $(y,s)=(-1,-\sqrt{1+1/k})$ in pseudo-tachyonic regime, then evolve
into the tachyonic regime towards a de Sitter expansion, represented by the
origin $(y,s)=(0,0)$. The trajectories of type IIa originate from the lines $%
s=\pm 1$ representing Big Bang singularities, then evolve from\ the
pseudo-tachyonic regime, through the corner points into the tachyonic era,
where they also reach the de Sitter attractor. The trajectories of type IIb
are similar, but they never cross into the pseudo-tachyonic regime

The trajectories of type IIIa and IIIb also originate in Big Bangs at the
lines $s=\pm 1$, pass through the tachyonic regime and through the
cornerpoint into a\ pseudo-tachyonic regime, followed by a collapse into a
Big Crunch (on the $y=\pm 1$ line). The difference between them is shown on
Fig. \ref{pszeudokvazi}. While IIIa reaches the soft singularity, evolves
through a short quasi-tachyonic regime in which the universe start to
recollapse towards a second soft singularity, those of type IIIb never reach
a soft singularity.

The trajectories of type IV originate in a Big Bang located in the
pseudo-tachyonic regime ($y=\pm 1$ line), evolve into the tachyonic regime,
then return into a pseudo-tachyonic regime, pass twice through soft
singularities (with a short quasi- tachyonic period), finally collapse into
a Big Crunch. The trajectories of type V are born in the Big Bang
singularities on the lines $s=\pm 1$, after which they mimic the evolution
of the trajectories of type IV following their second passage into a
pseudo-tachyonic regime.

On Fig. \ref{pszeudokvazi} we represent all types of trajectories which will
reach a pseudo-tachyonic regime in the future (types IIIa, IIIb, IV and V).
From among these IIIa, IV, V pass twice through soft singularities, with an
intercalated quasi-tachyonic regime, while those of type IIIb never reach a
soft singularity. We also determined numerically the critical evolutions
separating these two types of behaviors. They are the evolutions leading to
Big Brake type singularities in the presence of dust.

\section{Concluding Remarks}

In this paper we have expanded previous studies of a tachyonic cosmological
model with trigonometric potential and dust exhibiting a rather rich
spectrum of possible evolutions, by including curvature. As expected, a
positive curvature could counterbalance the influence of the dust
(manifesting itself in a positive $H$ and nonvanishing energy density at the
SFS), in the sense that it can compensate for the nonzero energy density at
the SFS. When the curvature term is strong enough, it stops the expansion
before the encounter with the SFS, preventing its formation. If the
curvature term is less strong, then the cosmological SFS is still formed,
followed by the transition to the quasi-tachyonic regime, but its
slowing-down effect is expected to reduce the duration of this regime as
compared to the flat case.

We have shown by numerical methods that it is possible to have a particular
balance between curvature and dust effects, leading to critical evolutions,
represented by the curves (4) and (6) on Fig. \ref{pszeudokvazi}. When such
a critical evolution is realized, the universe reaches exactly a Big Brake
singularity, which was forbidden in the flat case. The critical evolutions
reaching a traversable Big Brake separate evolutions into SFS from
completely regular ones. In this sense they resemble the criticality of
singularity-forming in gravitational collapse.\ This criticality is a new
feature of tachyonic cosmologies, induced by curvature. The possibility to
reach a Big Brake type singularity when dust is present is the second new
feature of the closed Friedmann universes with the tachyonic field.

\begin{acknowledgments}
We acknowledge stimulating interactions with David Polarski in the early
stages of this work. Z. K. was supported by OTKA Grant No. 100216 and A. K.
was partially supported by the RFBR Grant No. 14-02-00894.
\end{acknowledgments}

\end{document}